\documentclass[preprint,12pt, a4paper]{elsarticle}
\usepackage{amssymb, amsthm, amsmath}
\usepackage{graphicx}
\usepackage{multirow}
\usepackage{enumitem}
\usepackage{diagbox}
\usepackage{makecell}
\usepackage{float}
\usepackage{arydshln}
\usepackage{xcolor}
\usepackage{booktabs}
\usepackage{comment}
\usepackage{algorithm}
\usepackage{algpseudocode}
\usepackage{tikz}
\usepackage{listings}
\usepackage{xurl}
\usepackage{array}
\usepackage{hyperref}
\usetikzlibrary{shapes.geometric, arrows, positioning}
\theoremstyle{definition}

\newtheorem*{defn*}{Definition}
\theoremstyle{remark}

\newcommand{\Lc}{\mathcal{L}}
\newcommand{\Mc}{\mathcal{M}}

\newcommand{\Sc}{\mathcal{S}}

\def\1v{\mathbf 1}
\def\0v{\mathbf 0}

\begin{document}
\renewcommand{\labelenumii}{\arabic{enumi}.\arabic{enumii}}

\begin{frontmatter}

\title{iLBA: An R package for confidentially disseminating aggregated frequency tables}
\author[1]{Jeehyun Hwang\fnref{equal}}
\address[1]{Department of Statistics, Seoul National University, Seoul, Republic of Korea, hwangjh@snu.ac.kr}
\author[2]{Dongsun Yoon\fnref{equal}}
\address[2]{Department of Statistics, University of Michigan, Ann Arbor, Michigan, USA, dsyoon@umich.edu}
\author[3]{Sungkyu Jung\corref{cor1}}
\address[3]{Department of Statistics, Seoul National University, Seoul, Republic of Korea, sungkyu@snu.ac.kr, Corresponding author.}
\author[4]{Min-Jeong Park}
\address[4]{Statistical Standards Division, Statistics Policy Bureau, Statistics Korea, Republic of Korea, mj.park.stat@gmail.com}
\author[5]{Inkwon Yeo}
\address[5]{Department of Statistics, Sookmyung Women's University, Seoul, Republic of Korea, inkwon@sookmyung.ac.kr}

\fntext[equal]{These authors contributed equally to this work.}

\begin{abstract}
Statistical agencies frequently release frequency tables derived from microdata, but small frequency cells may lead to disclosure risks. We present \texttt{iLBA}, an open-source \textsf{R} package for confidential dissemination of aggregated frequency tables. The package implements the Information-Loss-Bounded Aggregation (iLBA) algorithm, which combines Small Cell Adjustment (SCA) at the finest level table with an aggregation procedure that introduces controlled ambiguity while bounding information loss. The software enables users to construct masked finest level tables, generate confidential aggregated tables for selected variables, and obtain masked frequencies for single-cell queries. By providing an accessible implementation of the iLBA method, the package facilitates reproducible and efficient disclosure control for tabular data derived from microdata.
\end{abstract}


\end{frontmatter}


\section*{Metadata}
\begin{table}[H]
\begin{tabular}{|p{1cm}|>{\raggedright\arraybackslash}p{6.5cm}|>{\raggedright\arraybackslash}p{6.5cm}|}
\hline
\textbf{Nr.} & \textbf{Code metadata description} & \textbf{Metadata} \\
\hline
C1 & Current code version & v1.0.0 \\
\hline
C2 & Permanent link to code/repository used for this code version & \url{https://github.com/SLTLab-SNU/iLBA_package} \\
\hline
C3  & Permanent link to Reproducible Capsule & \url{https://github.com/SLTLab-SNU/iLBA_package}\\
\hline
C4 & Legal Code License   & GPL-3 \\
\hline
C5 & Code versioning system used & GIT \\
\hline
C6 & Software code languages, tools, and services used & \textsf{R} ($\geq$ 3.5) \\
\hline
C7 & Compilation requirements, operating environments \& dependencies & 
Required packages: data.table, dplyr, magrittr\\
\hline
C8 & If available Link to developer documentation/manual & \url{https://github.com/SLTLab-SNU/iLBA_package/blob/main/iLBA_1.0.0.pdf}\\
\hline
C9 & Support email for questions & hwangjh@snu.ac.kr \\
\hline
\end{tabular}
\caption{Code metadata (mandatory)}
\label{codeMetadata} 
\end{table}

\section{Motivation and significance}

In response to expanding demand for public data from users of statistical agencies, ensuring the confidentiality in the release of detailed frequency tables has become an important task \citep{Chipperfield2016, Rinott2018}. Various frequency tables can be generated from a microdata set, which typically contains individual level records with demographic attributes (variables) and hierarchical classifications, such as geographic variables \citep{Shlomo2015} (province, county, and town) and industrial variables (sectors, industry groups, industries, and sub-industries \citep{MSCI_GICS}). When the microdata set is expressed as detailed frequency tables, they inevitably contain small frequency cells, whose counts for specific combinations of attributes of variables are less than a predefined threshold $K$ (Thresholds such as 3 or 5 are commonly used, depending on agency and context). These small frequency cells induce disclosure risks since they may allow an intruder to identify individuals in the population. This risk can be dealt with by ensuring $K$-anonymity, which requires that each released cell represents at least $K$ individuals \citep{Sweeney2002}.

Users of statistical agencies often request various combinations of variables according to their analytical needs. This leads to the generation of a massive number of frequency tables that vary significantly depending on the variable combinations and hierarchical levels used in their construction. For instance, given geographic hierarchical variables such as province, county, and town, a \textit{finest level table} is defined at the most granular level (e.g., town) with all demographic attributes included. A \textit{coarser level table} is subsequently obtained by summing cells that share the same unit at a higher geographic level.  
The primary challenge in releasing these tables lies in masking small frequency cells across both finest and coarser levels to ensure $K$-anonymity, especially since these tables are typically disseminated simultaneously. Furthermore, even if small cells are well masked in invididual tables, users may infer protected counts by differencing the multiple released tables \citep{Rinott2018, Shlomo2015, Shlomo2018}.

In this paper, we introduce \texttt{iLBA} \textsf{R} package, 
which implements the Information-Loss-Bounded Aggregation (iLBA) algorithm recently proposed by \cite{Park2024}. The package provides confidentially masked frequency tables of all requested combinations of variables, along with summaries of information loss, defined as the absolute difference between the original and masked values. 
While traditional
Small Cell Adjustments (SCA) \citep{Hundepool2012} ensure $K$-anonymity with bounded information loss in individual cells, their application in the aggregation process often results in excessive information loss and fails to maintain $K$-anonymity against differencing-based inference \citep{Park2024,Hundepool2026}. To address these issues, the \texttt{iLBA}  builds upon the SCA framework by introducing controlled \textit{ambiguity} into the aggregated cell counts. This mechanism prevents users from inferring exact values across the entire dissemination process while ensuring that the information loss remains strictly bounded.

The \texttt{iLBA} method addresses a fundamental challenge for national statistical agencies: producing protected frequency tables from hierarchical microdata while strictly controlling disclosure risk. Its practical efficacy is demonstrated by its integration into the Statistical Geographic Information Service Plus (SGIS+) \citep{SGIS}, the official data dissemination platform operated by Ministry of Data and Statistics, Republic of Korea. 
In this production environment, \texttt{iLBA} is utilized to securely release grid-level statistical tables while maintaining essential hierarchical consistency.
By providing an open-source implementation in \textsf{R}, this package allows for seamless integration into the analytical workflows of statistical offices and makes this methodology accessible to the global community. Given the widespread use of hierarchical statistical tables in official statistics, the \texttt{iLBA} package offers a practical tool for disclosure control in official data dissemination.

\subsection{Related methods and software}
Various methods and software tools have been developed to mitigate disclosure risks in tabular data. A pioneering tool in this field is $\tau$-Argus \citep{dewolf2003, tauargus2009}, many of whose functions were subsequently implemented in the \textsf{R} package \texttt{sdcTable} \citep{meindl2017}. This package protects tables through
suppression, resulting in masked tables that contain “NA” values \citep{Meindl2011}. When applied to hierarchical structures, the suppression-based approach often leads to substantial information loss.
More recently, the cell key method (CKM) was introduced and implemented in the \textsf{R} package \texttt{cellKey} \citep{meindl2025}. While CKM has been adopted by several national statistical offices (NSOs) to protect both frequency tables and continuous data \citep{Thompson2013}, it is not inherently designed to handle hierarchical key variables.
Although the CKM can be adapted for hierarchical structures \citep{Eurostat2025}, it remains unclear whether such adaptations ensure bounded information loss or consistently satisfy $K$-anonymity.

\subsection{The iLBA method}

Our dissemination framework involves the simultaneous release of the finest-level frequency table alonside all aggregated tables derived from it. In such tabular data, low-frequency cells pose a significant identity disclosure risk, as cells representing only a few individuals may facilitate re-identification when combined with external information
\citep{Shlomo2015}. Consequently, the primary objective of the confidentiality masking system is to ensure that $K$-anonymity is preserved across all released tables. 

A dataset satisfies $K$-anonymity if the information for any individual is indistinguishable from at least $K-1$ other individuals \citep{Sweeney2002}. For frequency tables, this requirement is interpreted as follows: a cell count $f$ satisfies $K$-anonymity if $f=0$, representing no individuals, or if $f \ge K$, representing at least $K$ indistinguishable individuals.
Conversely, $K$-anonymity is \textit{violated} if the released data allows users to deduce that the true count $f$ satisfies
$1 \le f \le K-1$. 

We illustrate the iLBA method by demonstrating the masking of both the finest-level table and its associated coarser-level tables, ensuring that $K$-anonymity is strictly preserved. We begin with the procedure for masking the finest-level table, using a synthetic microdata set as an example.
Table~\ref{synthetic} presents the synthetic microdata set $\Mc$, which includes three hierarchical variables and three key variables. The key variables consist of \texttt{gender}, \texttt{education}, and \texttt{age}, with 2, 9, and 18 categories, respectively. The hierarchical variables \texttt{LA1} (local area level 1), \texttt{LA2}, and \texttt{LA3} represent geographic units arranged in a nested structure through successive subdivisions, resulting in 1, 5, and 78 units, respectively. Reconstructed variables \texttt{L1}, \texttt{L2}, and \texttt{L3} represent these nested hierarchy. 
Higher hierarchical levels correspond to more aggregated (coarser) geographic units, while lower levels represent more detailed (finer) units. 



\begin{table}[ht]
\centering
\small
\caption{Synthetic microdata set $\Mc$ including three hierarchical variables and three key variables.}
\label{synthetic}
\begin{tabular}{c ccc c ccc ccc}
\toprule
\textbf{ID} 
& \multicolumn{3}{c}{\textbf{hierarchical variables}} 
& \multicolumn{3}{c}{\textbf{key variables}} 
& \multicolumn{3}{c}{\textbf{hierarchy levels}} \\
\cmidrule(lr){2-4} \cmidrule(lr){5-7} \cmidrule(lr){8-11}
&\textbf{LA1} & \textbf{LA2} & \textbf{LA3} 
& \textbf{gender} & \textbf{edu} & \textbf{age} 
& \textbf{L1} & \textbf{L2} & \textbf{L3} \\
\midrule
1       & 01 & 04 & 07 & 2 & 6 & 4 & 01 & 0104 & 010407 \\
2       & 01 & 04 & 02 & 1 & 4 & 7 & 01 & 0104 & 010407 \\
3       & 01 & 01 & 05 & 1 & 6 & 6 & 01 & 0101 & 010105 \\
\vdots  & \vdots & \vdots & \vdots & \vdots & \vdots & \vdots & \vdots & \vdots & \vdots \\
999998  & 01 & 03 & 11 & 2 & 1 & 2 & 01 & 0103 & 010311 \\
999999  & 01 & 02 & 07 & 1 & 3 & 3 & 01 & 0102 & 010207 \\
1000000 & 01 & 05 & 12 & 1 & 1 & 2 & 01 & 0105 & 010512 \\
\bottomrule
\end{tabular}
\end{table}

\paragraph{Masking the finest-level table}
The finest-level table derived from the raw microdata in Table~\ref{synthetic} is presented in Table~\ref{finest}, which comprises 25,272 rows. Due to the nested hierarchy, the number of valid geographic combinations is limited to 78, and the total row count reflects these units across all categories of \texttt{gender}, \texttt{edu}, and \texttt{age}. For brevity, Table~\ref{finest} displays only the first and last three rows. 

\begin{table}[!ht]
\centering
\caption{Finest-level table from the microdata $\Mc$. The raw frequency $f$ is masked to $\tilde{f}^{\text{SCA}}$.}
\label{finest}
\small
\begin{tabular}{lll l r r r r}
\toprule
\textbf{L1} & \textbf{L2} & \textbf{L3} 
& \textbf{gender} & \textbf{edu} & \textbf{age}
& $f$ & $\tilde{f}^{\text{SCA}}$ \\
\midrule
01 & 0101 & 010101 & 1 & 1  & 1 & 438 & 438 \\
01 & 0101 & 010101 & 1 & 1  & 2 & 164 & 164\\
01 & 0101 & 010101 & 1 & 1  & 3 & \textbf{0} & \textbf{0}\\
\addlinespace
\vdots & \vdots & \vdots & \vdots & \vdots & \vdots & \vdots & \vdots \\
\addlinespace
01 & 0105 & 010512 & 2 & 9 & 16 & \textbf{1} & \textbf{5}\\
01 & 0105 & 010512 & 2 & 9 & 17 & \textbf{3} & \textbf{0}\\
01 & 0105 & 010512 & 2 & 9 & 18 & \textbf{5} & \textbf{5}\\
\bottomrule
\end{tabular}
\end{table}

We apply the SCA method to mask small frequency cells in Table~\ref{finest}, defined as those with counts below a predefined threshold $K$. 
The SCA replaces the true cell frequency $f$ with its masked value $\tilde{f}^{\text{SCA}}$ as follows. 
\begin{equation}
\tilde{f}^{\text{SCA}} =
\begin{cases}
f^*, & f \in \{0, \ldots, K\}, \\[4pt]
f, & f \in \{K+1, \ldots\},
\end{cases}
\tag{1}
\label{1}
\end{equation}
\noindent where the value of $f^*$ is given at random among $\{0,K\}$: 
\[
f^* =
\begin{cases}
0, & \text{with probability } 1 - \dfrac{f}{K}, \\[6pt]
K, & \text{with probability } \dfrac{f}{K}.
\end{cases}
\]
The SCA leaves a cell unchanged when its count is at least $K$. Since $|\tilde{f}^{\text{SCA}}-f|\leq K-1$, the SCA gaurantees bounded information loss and ensures $K$-anonymity, because all released cell counts are $0$ or at least $K$. For the illustration in Table~\ref{finest}, we set $K=5$. 

\paragraph{Masking coaser-level tables} The subsequent phase involves masking aggregated, coarser-level tables while maintaining $K$-anonymity. To illustrate this procedure, consider a hypothetical user request for an aggregated count where (\texttt{L3}, \texttt{gender}, \texttt{edu}) = (010101, 2, 2). The corresponding 18 cells are extracted from Table~\ref{finest} to form the subset presented in Table~\ref{selected cells}. 

\begin{table}[t]
\centering
\caption{Selected cells from Table~\ref{finest} according to the user's request (\texttt{L3}, \texttt{gender}, \texttt{edu}) = (010101, 2, 2).}
\label{selected cells}
\small
\begin{tabular}{cccc ccc cc}
\toprule
\textbf{j} & \textbf{L1} & \textbf{L2} & \textbf{L3} 
& \textbf{gender} & \textbf{edu} & \textbf{age}
& $f_j$ & $\tilde{f}_j^{\text{SCA}}$ \\
\midrule
1 & 01 & 0101 & 010101 & 2 & 2 & 1  & 36   & 36   \\
2 & 01 & 0101 & 010101 & 2 & 2 & 2  & 284 & 284 \\
3 &01 & 0101 & 010101 & 2 & 2 & 3  & 262 & 262 \\
4 &01 & 0101 & 010101 & 2 & 2 & 4  & \textbf{1}   & \textbf{5}   \\
5 &01 & 0101 & 010101 & 2 & 2 & 5  & \textbf{1}   & \textbf{5}  \\
6 &01 & 0101 & 010101 & 2 & 2 & 6  & \textbf{2}   & \textbf{5}  \\
7 & 01 & 0101 & 010101 & 2 & 2 & 7  & \textbf{1}   & \textbf{5}   \\
8 & 01 & 0101 & 010101 & 2 & 2 & 8  & \textbf{1}   & \textbf{0}  \\
9 & 01 & 0101 & 010101 & 2 & 2 & 9  & 10  & 10  \\
10 & 01 & 0101 & 010101 & 2 & 2 & 10 & 9   & 9   \\
11 & 01 & 0101 & 010101 & 2 & 2 & 11 & 79  & 79  \\
12 & 01 & 0101 & 010101 & 2 & 2 & 12 & 124 & 124 \\
13 & 01 & 0101 & 010101 & 2 & 2 & 13 & 130 & 130 \\
14 & 01 & 0101 & 010101 & 2 & 2 & 14 & 106 & 106 \\
15 & 01 & 0101 & 010101 & 2 & 2 & 15 & 125 & 125  \\
16 & 01 & 0101 & 010101 & 2 & 2 & 16 & 77  & 77  \\
17 & 01 & 0101 & 010101 & 2 & 2 & 17 & 60  & 60  \\
18 & 01 & 0101 & 010101 & 2 & 2 & 18 & 18  & 18  \\
\midrule
& & & & & & & $f = 1326\quad (f_{\Sc}= 6)$\\
\bottomrule
\end{tabular}
\end{table}

To formalize the iLBA algorithm, the subset of cells to be aggregated is partitioned based on their masked values. 
By indexing these cells as $[m] = \{1, \dots, m\}$, we identify the indices of the finest-level cells whose SCA-masked values are $0$ or $K$, respectively: 
$$
\Sc_0 = \{j \in [m]: \tilde{f}_j^{\text{SCA}} = 0\}, \, \, 
\Sc_K = \{j \in [m]: \tilde{f}_j^{\text{SCA}} = K\}, \, \, 
\Sc = \Sc_0 \cup \Sc_K.
$$
The set $\Sc$ thus represents the collection of ``small cells'' where $f_j \le K$. 
The total aggregated count $f = \sum_{j \in [m]} f_j$ is then decomposed into contributions from both small and large cells: 
$$
f_\Sc = \sum_{j: f_j \le K} f_j, \, \,
f_\Lc = \sum_{j: f_j > K} f_j, \, \, 
f = f_\Sc + f_\Lc.$$
Applying these definitions to the example in Table~\ref{selected cells} (where $\Sc=\{4,\dots,8\}$, $\Sc_0 = \{8\}$, and $\Sc_K = \{4,5,6,7\}$), we obtain $f_{\Sc} = 6$ and $f_{\Lc} = 1320$.

Since $f_{\Lc}$ is known exactly from the finest-level table, the security of the aggregated count $f$ depends entirely on the masking of $f_{\Sc}$. Naive approaches to mask $f_{\Sc}$ are often inadequate. For instance, releasing the sum of individual SCA-masked counts, $\tilde{f}_{\Sc}^{\text{sum}} = \sum_{j \in \Sc} \tilde{f}_j^{\text{SCA}}$, results in excessive information loss (in our example, $|\tilde{f}_{\Sc}^{\text{sum}} - f_{\Sc}| = 14$). Alternatively, applying the SCA rule directly to $f_{\Sc}$ might leave the true value unchanged (e.g., $\tilde{f}_{\Sc}^{\text{SCA}} = 6$), revealing $f_{\Sc}$ precisely. Releasing such information may allow users to infer the underlying small frequency cells in the finest-level table. Such inference, achieved by differencing the released tables, violates $K$-anonymity; see \ref{pitfalls}. 

 
To mitigate this risk, 
the aggregated output must retain sufficient uncertainty. We formalize this requirement as \textit{$K$-ambiguity}: a masked count $\tilde{f}$ satisfies this condition if at least $K$ candidate true values are compatible with the released information.  
The \texttt{iLBA} algorithm is specifically designed to fulfill this dual requirement: achieving $K$-anonymity across all released tables while employing $K$-ambiguity as an aggregation-level safeguard against differencing-based inference. 

As detailed in Algorithm~\ref{alg:iLBA}, the \texttt{iLBA} aggregation procedure takes the true aggregated count $f_\Sc$ and the numbers of small cells masked to $0$ and $K$ (denoted as $|\Sc_0|$ and $|\Sc_K|$) as primary inputs. The algorithm first constructs an initial candidate set $C$ of length $K$, ensuring that $C$ contains the true frequency $f_{\Sc}$. To maintain statistical plausibility, the procedure evaluates whether $C$ lies within the feasible interval $D$, the range of possible sums constrained by the SCA-masked small cells. If $C$ falls outside this range, the algorithm shifts the set to ensure that $K$-ambiguity is strictly satisfied. 
A post-processing rule is then applied to ensure the masked small-cell sum $\tilde{f}_{\Sc}^{\text{iLBA}}$ is either $0$ or at least $K$, preserving $K$-anonymity at the aggregated level. The final released count is computed as $\tilde{f} = \tilde{f}_{\Sc}^{\text{iLBA}} + f_{\Lc}.$
This value $\tilde{f}$ is provided to users in place of the true aggregated count $f$. In the example from Table~\ref{selected cells}, $\tilde{f}_{\Sc}^{\text{iLBA}} = 8$ and $\tilde{f} = 1328$, resulting in a minimal information loss of   $|\tilde{f}-f|=2$.  

A step-by-step breakdown of Algorithm~\ref{alg:iLBA} is provided in \ref{algorithm detail}.

\begin{algorithm}[tp!]
\caption{Loss-Bounded Aggregation (iLBA)}
\label{alg:iLBA}
\begin{algorithmic}[1] 
\renewcommand{\algorithmicrequire}{\textbf{Input:}}
\renewcommand{\algorithmicensure}{\textbf{Output:}}

\Require Small-cell index sets $\Sc_0$ and $\Sc_K$, true aggregated count $f_{\Sc}$, threshold $K$
\Ensure Masked value $\tilde{f}_{\Sc}^{\text{iLBA}}$
\Statex 

\If{$|\Sc| = 0$ \textbf{or} $f_{\Sc} = 0$} 
    \State \Return $\tilde{f}_{\Sc}^{\text{iLBA}} = 0$
\ElsIf{$\Sc = \{j_0\}$ for some $j_0$}  
    \State \Return $\tilde{f}_{\Sc}^{\text{iLBA}} = \tilde{f}_{j_0}^{\text{SCA}} \in \{0, K\}$
\Else 
    \State \textbf{step 1: Compute initial candidate center:}
    \State $\tilde{f}_{\Sc}^{(1)} \gets f_{\Sc} - \text{mod}(f_{\Sc} - 1, K) + \lfloor K/2 \rfloor$
    
    \State \textbf{step 2: Define candidate set $C$:}
    \State $C \gets \{ \tilde{f}_{\Sc}^{(1)} - \lfloor K/2 \rfloor, \dots, \tilde{f}_{\Sc}^{(1)} - \lfloor K/2 \rfloor + K - 1 \}$
    
    \State \textbf{step 3: Adjust for feasible interval $D$ (Shifting):}
    \State $D \gets \Big\{|\Sc_K|,|\Sc_{K}|+1,\dots,\; K|\Sc_K| + (K-1)|\Sc_0|\Big\}$
    \If{$\min(C) < \min(D)$}
        \State $\tilde{f}_{\Sc}^{(2)} \gets \tilde{f}_{\Sc}^{(1)} + K$
    \ElsIf{$\max(D) < \max(C)$}
        \State $\tilde{f}_{\Sc}^{(2)} \gets \tilde{f}_{\Sc}^{(1)} - K$
    \Else
        \State $\tilde{f}_{\Sc}^{(2)} \gets \tilde{f}_{\Sc}^{(1)}$
    \EndIf
    
    \State \textbf{step 4: Apply post-processing rule:}
    \State $\tilde{f}_{\Sc}^{(3)} \gets \begin{cases} K, & \text{if } \tilde{f}_{\Sc}^{(2)} = 1 + \lfloor K/2 \rfloor \\ \tilde{f}_{\Sc}^{(2)}, & \text{otherwise} \end{cases}$
    
    \State \Return $\tilde{f}_{\Sc}^{\text{iLBA}} = \tilde{f}_{\Sc}^{(3)}$
\EndIf

\end{algorithmic}
\end{algorithm}

\paragraph{Guarantees of the iLBA algorithm}
For a fixed threshold $K \ge 3$, the following properties hold:
\begin{enumerate}
\item (\emph{Bounded information loss}) The absolute information loss is bounded:
      \begin{equation*}
          |\tilde{f}-f| \leq
        \begin{cases}
            \lfloor K/2 \rfloor + K, &\text{if} \quad|\Sc|\geq2 \quad \text{and} \quad f_{\Sc}\geq1,\\
            K-1, &\text{otherwise}.
        \end{cases}
      \end{equation*}
      Note that when no shift is applied in Step 3 of Algorithm~\ref{alg:iLBA} and the post-processing in Step 4 is not triggered (equivalently, $f^{\text{(1)}}_{\Sc} = f^{\text{(2)}}_{\Sc}$ and $f^{\text{(2)}}_{\Sc} \neq 1+ \lfloor K/2 \rfloor$), the information loss is $|\tilde{f}-f| \le \lfloor K/2 \rfloor$, which generates very small information loss.
\item (\emph{$K$-ambiguity}) The released value $\tilde f^{\mathrm{iLBA}}_{\Sc}$ ensures $K$-ambiguity.

\item  (\emph{$K$-anonymity at both levels})
By construction, the released count $\tilde f^{\mathrm{iLBA}}_{\Sc}$ is either $0$ or at
least $K$, so every aggregated count satisfies
$K$-anonymity. Moreover, the $K$-ambiguity of $\tilde f^{\mathrm{iLBA}}_{\Sc}$ guarantees that users cannot uniquely assign any individual finest-level cell in $\Sc$ a specific true count within the sensitive range $\{1, \dots, K-1\}$, even when combined with the known SCA rules. Consequently, $K$-anonymity is preserved for both the aggregated and the finest-level counts. 
A formal mathematical proof of how $K$-ambiguity prevents such disclosure is provided in \ref{K-ambiguity}.
\end{enumerate}

\section{Software description}
 The \texttt{iLBA} \textsf{R} package is designed to enable users to obtain confidentially masked tables and frequencies from microdata. The source code of the package is available at \url{https://github.com/SLTLab-SNU/iLBA_package}. The package can be installed from the \texttt{R} console using the following commands.

\begin{verbatim}
> install.packages("remotes")
> remotes::install_github("SLTLab-SNU/iLBA_package")
> library(iLBA)
\end{verbatim}

\subsection{Software architecture}

\tikzset{
  common/.style={
    font=\small,
    node distance=1.2cm and 3.0cm
  },
  block/.style={
    draw, rectangle,
    minimum width=3.2cm,
    minimum height=0.8cm,
    align=center,
    font=\small
  },
  title/.style={font=\bfseries\small},
  arrow/.style={->, >=Stealth, thick}
}

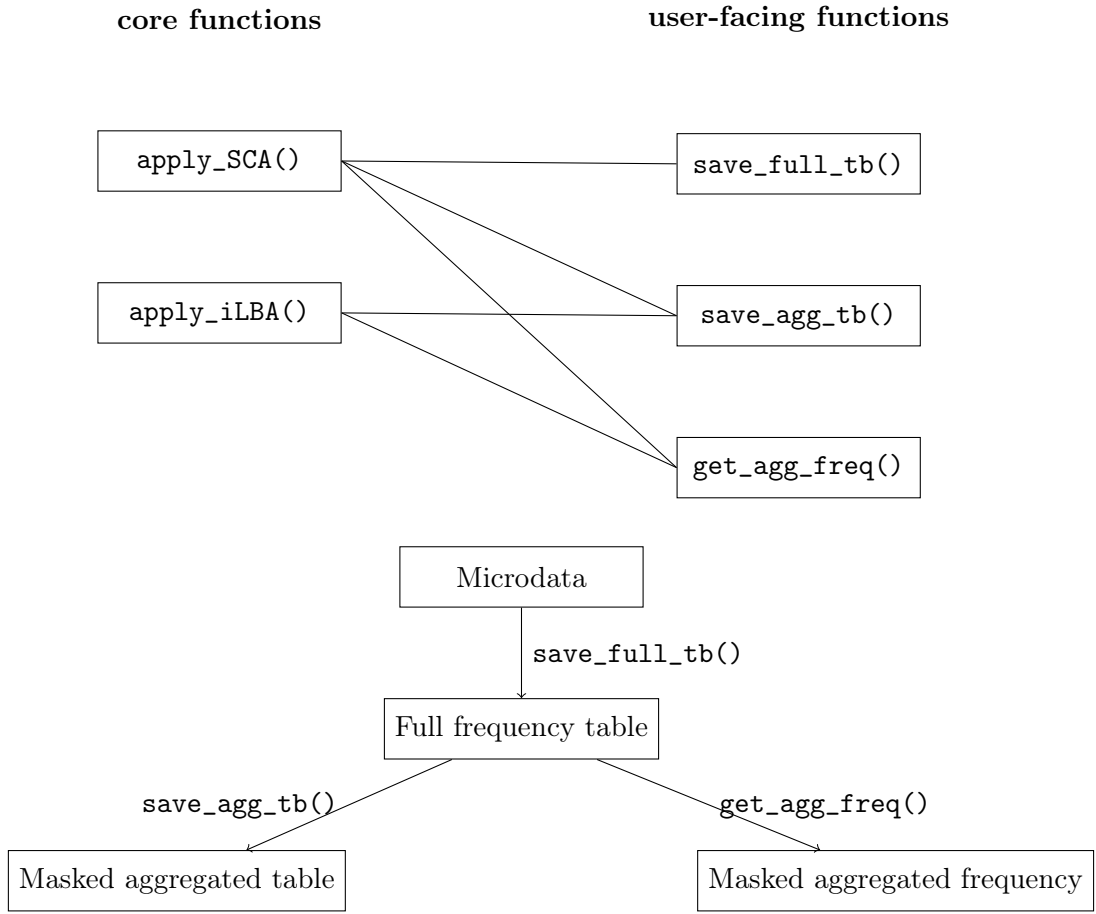
\begin{figure}[t] 
\centering

\begin{tikzpicture}[common]
  \node[title] (core_title) {core functions};
  \node[title, right=of core_title, xshift=1.0cm] (user_title) {user-facing functions};

  \node[block, below=of core_title] (sca) {\texttt{apply\_SCA()}};
  \node[block, below=of sca] (ilba) {\texttt{apply\_iLBA()}};

  \node[block, below=of user_title] (full_tb) {\texttt{save\_full\_tb()}};
  \node[block, below=of full_tb] (agg_tb) {\texttt{save\_agg\_tb()}};
  \node[block, below=of agg_tb] (agg_freq) {\texttt{get\_agg\_freq()}};

  \draw (sca.east) -- (full_tb.west);
  \draw (sca.east) -- (agg_tb.west);
  \draw (sca.east) -- (agg_freq.west);
  \draw (ilba.east) -- (agg_tb.west);
  \draw (ilba.east) -- (agg_freq.west);
\end{tikzpicture}

\vspace{0.6cm} 

\begin{tikzpicture}[common, node distance=1.2cm and 0.5cm]
  \node[block] (micro) {Microdata};
  \node[block, below=of micro] (full) {Full frequency table};

  \node[block, below left=of full] (masked_agg) {Masked aggregated table};
  \node[block, below right=of full] (masked_freq) {Masked aggregated frequency};

  \draw[->] (micro) -- node[right] {\texttt{save\_full\_tb()}} (full);
  \draw[->] (full) -- node[left] {\texttt{save\_agg\_tb()}} (masked_agg);
  \draw[->] (full) -- node[right] {\texttt{get\_agg\_freq()}} (masked_freq);
\end{tikzpicture}

\caption{(Top) Two-layer software architecture. (Bottom) Workflow from microdata to masked outputs.}
\label{flowchart}
\end{figure}

The \texttt{iLBA} \textsf{R} package is built in two layers: core masking functions and user-facing workflow
functions. The \textit{core layer} consists of \texttt{apply\_SCA()} and \texttt{apply\_iLBA()}, which implement
the the privacy-preserving masking procedures defined in \eqref{1} and Algorithm~\ref{alg:iLBA}, respectively. 
The \textit{user-facing layer} provides high-level functions---\texttt{save\_\allowbreak full\_\allowbreak tb()},
\texttt{save\_\allowbreak agg\_\allowbreak tb()} and \texttt{get\_\allowbreak agg\_\allowbreak freq()}---that manage the data pipeline from raw microdata to masked tabular outputs.
Figure~\ref{flowchart} illustrates this two-layer architecture and overall workflow of the package. 

Given a microdata set, \texttt{save\_\allowbreak full\_\allowbreak tb()} first constructs the finest level table and applies \texttt{apply\_\allowbreak SCA()} to each observed cell count. For computational efficiency, only observed combinations of variables are written in the finest level table, whereas zero count combinations are omitted. This design substantially reduces storage and computation, while leaving subsequent aggregation results unchanged because omitted combinations contribute zero to any aggregated count. 

The stored finest level table generated by \texttt{save\_\allowbreak full\_\allowbreak tb()} is then used in two ways. First, \texttt{save\_\allowbreak agg\_\allowbreak tb()} produces masked coarser level tables for user-selected hierarchical levels and key variables. 
Conceptually, at the requested hierarchical level, it groups the cells of the finest level table according to all combinations of the selected key variables, aggregates over lower level hierarchical units and omitted key variables, and then applies \texttt{apply\_iLBA()} to each aggregated cell. 
Second, \texttt{get\_\allowbreak agg\_\allowbreak freq()} returns the masked frequency for a single target cell defined by a user-specified set of variable--attribute pairs. 
At the requested hierarchical level, it extracts the finest level cells corresponding to that target cell, aggregates their counts, and applies \texttt{apply\_iLBA()} once to the aggregated count. 
Thus, while \texttt{save\_\allowbreak agg\_\allowbreak tb()} applies \texttt{apply\_iLBA()} repeatedly across all aggregated cells, \texttt{get\_\allowbreak agg\_\allowbreak freq()} applies it only once for the requested cell. 
Because both functions rely on the same masking procedure, the value returned by \texttt{get\_\allowbreak agg\_\allowbreak freq()} is consistent with the corresponding entry in the aggregated tables produced by \texttt{save\_\allowbreak agg\_\allowbreak tb()}.

\subsection{Software functionalities}
The main user-facing functions of the package are \texttt{save\_\allowbreak full\_\allowbreak tb()}, \texttt{save\_\allowbreak agg\_\allowbreak tb()}, and \texttt{get\_\allowbreak agg\_\allowbreak freq()}.

\begin{verbatim}
save_full_tb( 
    data, 
    hkey, 
    key = NULL, 
    mask_thr = 5, 
    hkey_rank = NULL, 
    key_thr = 100, 
    output_path = "full_tb.rds")
\end{verbatim}

The function \texttt{save\_\allowbreak full\_\allowbreak tb()} is the entry point for constructing the finest level frequency table from a microdata set. The user supplies a \texttt{data.frame} or \texttt{data.table}, the hierarchical variables (\texttt{hkey}), and optionally the key variables (\texttt{key}). If \texttt{key} is omitted, all non-hierarchical variables are used. The function requires at least one hierarchical variable. However, it can still be applied to datasets containing only key variables by designating one of the key variables as a hierarchical variable. The hierarchical variables should be specified either from coarser to finer levels or together with an optional argument \texttt{hkey\_rank}. If \texttt{hkey} is not ordered from coarser to finer levels, \texttt{hkey\_rank} must be provided as a vector of the same length indicating the hierarchical rank of each variable (e.g., province: 1, county: 2, town: 3). To avoid including quantitative variables, the function can exclude key variables whose number of categories exceeds a user-specified threshold \texttt{key\_thr}, which defaults to 100. The function then applies \texttt{apply\_SCA()} using the threshold \texttt{mask\_thr}, which defaults to $K=5$, and saves an RDS object in \texttt{output\_path}, containing the finest level table, masked counts, and metadata such as variable names and category sets. The function also produces console output displaying a list of the hierarchical variables with their ranks, a list of the key variables, the masking threshold, and the output file path. This console output helps users specify inputs for subsequent functions.

\begin{verbatim}
save_agg_tb( 
    hkey_level, 
    key, 
    input_path = "full_tb.rds", 
    output_tb_path = "agg_tb.csv", 
    output_iL_path = "info_loss.csv")
\end{verbatim}

The function \texttt{save\_\allowbreak agg\_\allowbreak tb()} generates a masked coarser level table from a previously saved finest-level table. The user specifies the target hierarchical level (\texttt{hkey\_level}), the key variables to select (\texttt{key}), and the path to the RDS object (\texttt{input\_path}) produced by \texttt{save\_\allowbreak full\_\allowbreak tb()}. The hierarchical level must be provided as an integer and can be identified easily from the console output of \texttt{save\_\allowbreak full\_\allowbreak tb()}. For datasets with a single hierarchical variable, the level should be specified as $1$. 
The function computes the true aggregated counts for all combinations of selected variables at the requested hierarchical level, applies \texttt{apply\_iLBA()} to each aggregated cell, and writes the resulting masked table to a CSV file at the user-specified \texttt{output\_tb\_path}. In addition, a CSV file summarizing the differences between the true and masked counts is saved at \texttt{output\_iL\_path}.

\begin{verbatim}
get_agg_freq( 
    hkey_level, 
    key, 
    hkey_value, 
    key_value, 
    input_path = "full_tb.rds")
\end{verbatim}

The function \texttt{save\_\allowbreak agg\_\allowbreak tb()} returns a masked frequency for a user-specified cell. The user provides the hierarchical level (\texttt{hkey\_level}) as an integer, the key variables to select (\texttt{key}), the corresponding hierarchical and key values (\texttt{hkey\_value} and \texttt{key\_value}) that define the target cell, and the path to the stored finest-level table (\texttt{input\_path}). Internally, the function extracts the cells from the finest level table constituting the target cell. It then sums their counts, applies \texttt{apply\_iLBA()} to the aggregated count, and returns the masked frequency. This function is useful when a user needs a protected value for a specific cell without generating the full aggregated table.

\section{Illustrative examples}

\subsection{Census Dataset}

Table~\ref{census} shows a synthetic census dataset, which is included in the package for illustration and
analysis. The dataset contains 1,000,000 records, four hierarchical
key variables (\texttt{LA1}--\texttt{LA3} and \texttt{OA}) and five key variables (\texttt{gender}, \texttt{age}, \texttt{edu}, \texttt{mar}, and \texttt{htype}).
\texttt{LA1}--\texttt{LA3} and \texttt{OA} denote geographic units in a nested hierarchy: \texttt{LA2} subdivides \texttt{LA1}, \texttt{LA3}
subdivides \texttt{LA2}, and \texttt{OA} (Output Area) represents the smallest statistical area unit. In this dataset, synthetic data generation was used to replace private personal information mimicking the distribution of the original 2010 Census microdata of Korea. The original data is available at the Statistics Data Center (SDC) at the Ministry of Data and Statistics (MODS) \citep{SDC} in a secure environment. The census dataset can be loaded and viewed in \textsf{R} by using the following commands.
\begin{verbatim}
#Load the package
library(iLBA)
#Load data
data(census)
#View the first few rows
head(census)
\end{verbatim}

\begin{table}[!htbp]
\centering
\caption{Census dataset. The numbers of categories are 1 (LA1), 5 (LA2), 78 (LA3), and
2506 (OA) for hierarchical variables, and 2 (gender), 18 (age), 9 (edu), 5 (mar), and 21
(htype) for key variables.}
\label{census}
\centering
\begin{tabular}[t]{ccccccccc}
\toprule
LA1 & LA2 & LA3 & OA & gender & age & edu & mar & htype\\
\midrule
01 & 0104 & 010407 & 01040704 & 2 & 4 & 6 & 1 & 21\\
01 & 0104 & 010402 & 01040237 & 1 & 7 & 4 & 1 & 19\\
01 & 0101 & 010105 & 01010504 & 1 & 6 & 6 & 1 & 21\\
01 & 0101 & 010108 & 01010815 & 2 & 4 & 6 & 1 & 28\\
01 & 0104 & 010403 & 01040346 & 2 & 10 & 3 & 2 & 33\\
\addlinespace
\vdots & \vdots & \vdots & \vdots & \vdots & \vdots & \vdots & \vdots  & \vdots\\
\addlinespace
01 & 0104 & 010406 & 01040648 & 2 & 4 & 6 & 1 & 99\\
01 & 0102 & 010212 & 01021201 & 1 & 7 & 6 & 2 & 21\\
01 & 0103 & 010310 & 01031013 & 1 & 9 & 8 & 4 & 22\\
01 & 0105 & 010512 & 01051246 & 2 & 13 & 2 & 2 & 21\\
01 & 0101 & 010104 & 01010434 & 2 & 3 & 3 & 9 & 21\\
\bottomrule
\end{tabular}
\end{table}

\subsection{Construct the finest level table}

Suppose a statistical agency has just completed a population census and intends to disseminate frequency tables. The agency’s objective is to release these tables in a confidential manner. The first step for the agency is to call \texttt{save\_\allowbreak full\_\allowbreak tb()} with the appropriate hierarchical key variables and key variables. Here, we use all variables included in the census dataset. For the \texttt{hkey} input, the agency should specify hierarchical variables either in the descending hierarchical order or in arbitrary order with \texttt{hkey\_rank} option (e.g., \texttt{hkey = c("LA2",\allowbreak "LA1",\allowbreak "OA",\allowbreak "LA3"),\allowbreak hkey\_rank = \allowbreak c(2,1,4,3)}). The function \texttt{save\_\allowbreak full\_\allowbreak tb()} constructs the finest level frequency table and applies the SCA to each cell count. Table~\ref{census_finest} is the resulting table that contains both true and masked values. The table is saved as an RDS object at the specified output path.

\begin{verbatim}
save_full_tb(
    data = census,
    hkey = c("LA1","LA2","LA3","OA"),
    key = c("gender", "age", "edu", "mar", "htype"),
    mask_thr = 5,
    output_path = "full_tb.rds"
)
\end{verbatim}

\begin{table}[!htbp]
\centering
\caption{The SCA masked finest level frequency table}
\label{census_finest}
\centering
\begin{tabular}[t]{ccccccccccc}
\toprule
LA1 & LA2 & LA3 & OA & gender & age & edu & mar & htype & N & N\_masked\\
\midrule
01 & 0104 & 010407 & 01040704 & 2 & 4 & 6 & 1 & 21 & 3 & 5\\
01 & 0104 & 010402 & 01040237 & 1 & 7 & 4 & 1 & 19 & 1 & 0\\
01 & 0101 & 010105 & 01010504 & 1 & 6 & 6 & 1 & 21 & 2 & 0\\
01 & 0101 & 010108 & 01010815 & 2 & 4 & 6 & 1 & 28 & 1 & 0\\
01 & 0104 & 010403 & 01040346 & 2 & 10 & 3 & 2 & 33 & 1 & 5\\
\addlinespace
\vdots & \vdots & \vdots & \vdots & \vdots & \vdots & \vdots & \vdots & \vdots & \vdots & \vdots\\
\addlinespace
01 & 0104 & 010406 & 01040648 & 2 & 4 & 6 & 1 & 99 & 1 & 0\\
01 & 0102 & 010212 & 01021201 & 1 & 7 & 6 & 2 & 21 & 6 & 6\\
01 & 0103 & 010310 & 01031013 & 1 & 9 & 8 & 4 & 22 & 1 & 5\\
01 & 0105 & 010512 & 01051246 & 2 & 13 & 2 & 2 & 21 & 2 & 0\\
01 & 0101 & 010104 & 01010434 & 2 & 3 & 3 & 9 & 21 & 4 & 5\\
\bottomrule
\end{tabular}
\end{table}

\subsection{Aggregate at a coarser level with iLBA masking}

Now, a user can request frequency tables at multiple geographic levels and for various combinations of key variables. Suppose the user wants to obtain a table at the third geographic level (\texttt{LA3}) using only \texttt{gender}, \texttt{age} and \texttt{htype} key variables. Since the hierarchical order of the finest level table is specified as \texttt{LA1}, \texttt{LA2}, \texttt{LA3} and \texttt{OA} when executing \texttt{save\_\allowbreak full\_\allowbreak tb()}, the input \verb|hkey_level| of \texttt{save\_\allowbreak agg\_\allowbreak tb()} for \texttt{LA3} is 3.  The function outputs two CSV files: (i) the masked aggregated table and (ii) the corresponding information-loss summary. Figure~\ref{census_console} shows the console output produced when the code is executed. 

\begin{verbatim}
save_agg_tb(
    hkey_level = 3,
    key = c("gender","age","htype"),
    input_path = "full_tb.rds",
    output_tb_path = "agg_tb.csv",
    output_iL_path = "info_loss.csv"
)
\end{verbatim}

\begin{figure}[tbp!]
\centering
\begin{minipage}{0.9\textwidth}
\begin{lstlisting}[basicstyle=\ttfamily\footnotesize]
Header of aggregated masked table via iLBA
      LA1   LA2   LA3   gender age htype N_masked type1 type2
      <char> <char> <char> <int> <int> <int> <int> <int> <int>
      01    0101  010101 1     1   21    315      0     0
      01    0101  010101 1     1   22    8        0     0
      01    0101  010101 1     1   23    18       0     0
      01    0101  010101 1     1   26    8        0     0
      01    0101  010101 1     1   27    0        0     0
      01    0101  010101 1     1   29    13       0     0

Distribution of Information Loss
      Loss   n     perc
      -4     1     0.00
      -3     1     0.00
      -2     3489  9.48
      -1     8780  23.86
       0     5168  14.04
       1     6031  16.39
       2     7009  19.05
       3     4173  11.34
       4     1713  4.65
       5     283   0.77
       6     154   0.42
      Total 36802 100.00
\end{lstlisting}
\end{minipage}
\caption{The coarser level table and its information loss summary.}
\label{census_console}
\end{figure}

In practice, statistical agencies typically fix the set of key variables to be released and run \texttt{save\_\allowbreak agg\_\allowbreak tb()} once for each hierarchical geographic level. After generating these masked aggregated tables, the agency can store them and directly use them for public dissemination.


\subsection{Computational performance}

We evaluated the computational performance of \texttt{save\_\allowbreak full\_\allowbreak tb()} and \texttt{save\_agg\_tb()} using the census dataset. For \texttt{save\_\allowbreak full\_\allowbreak tb()}, we generated the finest level table with \texttt{hkey = c("LA1",\allowbreak"LA2",\allowbreak"LA3",\allowbreak"OA")} and \texttt{key = c("gender",\allowbreak"age",\allowbreak"edu",\allowbreak"mar",\allowbreak"htype")}. This computation completed in 1.50 s and produced the finest level table with 617,543 nonzero rows. Here, the number of rows refers to the number of observed nonzero combinations of area units and key variable attributes that actually appear in the dataset, rather than the full cartesian product of all possible combinations. For the finest level table, the full cartesian product of variables is \(2506 \times (2 \times 18 \times 9 \times 5 \times 21) = 85{,}254{,}120\), but only a small fraction of these combinations are observed in the dataset.

We further benchmarked \texttt{save\_\allowbreak agg\_\allowbreak tb()} by varying the hierarchical level and the number of key variables from one to five (see Table~\ref{experiment_results}). The results demonstrate that while the runtime fluctuates slightly for outputs of smaller rows, the overall execution time is strongly driven by the number of generated nonzero rows. That is, adding more key variables affects runtime primarily when it substantially increases the size of the output table. Consequently, computations remain fast at higher hierarchical levels (i.e., closer to 1), but require more time at lower hierarchical levels (i.e., closer to 4) where significantly more combinations of variables must be processed.

\begin{table}[h]
\centering
\caption{Runtime and output table size of \texttt{save\_agg\_tb()} by hierarchical level and number of key variables}
\label{experiment_results}
\begin{tabular}{ccp{5.8cm}cc}
\hline
hkey level & \# keys & keys used & Time (sec) & \# rows \\
\hline
1 & 1 & gender & 0.3234 & 2 \\
1 & 2 & gender, mar & 0.3210 & 10 \\
1 & 3 & gender, mar, edu & 0.3262 & 79 \\
1 & 4 & gender, mar, edu, age & 0.3209 & 777 \\
1 & 5 & gender, mar, edu, age, htype & 0.3761 & 5140 \\
\hline
2 & 1 & gender & 0.3273 & 10 \\
2 & 2 & gender, mar & 0.3095 & 50 \\
2 & 3 & gender, mar, edu & 0.3275 & 387 \\
2 & 4 & gender, mar, edu, age & 0.3536 & 3591 \\
2 & 5 & gender, mar, edu, age, htype & 0.5905 & 21474 \\
\hline
3 & 1 & gender & 0.4225 & 156 \\
3 & 2 & gender, mar & 0.3223 & 780 \\
3 & 3 & gender, mar, edu & 0.3916 & 5627 \\
3 & 4 & gender, mar, edu, age & 0.9399 & 37070 \\
3 & 5 & gender, mar, edu, age, htype & 2.2061 & 145061 \\
\hline
4 & 1 & gender & 0.3697 & 5012 \\
4 & 2 & gender, mar & 0.6516 & 24777 \\
4 & 3 & gender, mar, edu & 1.8744 & 116297 \\
4 & 4 & gender, mar, edu, age & 5.5057 & 370774 \\
4 & 5 & gender, mar, edu, age, htype & 9.3197 & 617543 \\
\hline
\end{tabular}
\end{table}

\section{Impact and conclusions}

The Statistical Geographic Information Service Plus (SGIS+) is a user-friendly data dissemination platform of Ministry of Data and Statistics, Repulbic of Korea, that provides official statistics through interactive, map-based interfaces. It allows users to generate and visualize frequency tables across multiple administrative areas or at various grid levels, enabling detailed statistical exploration at different regional levels. Within this system, the \texttt{iLBA} algorithm was implemented in Java to integrate with the platform’s Java-based infrastructure in 2021. The \texttt{iLBA} algorithm is currently used to disseminate statistics from multiple national surveys, including the Population and Housing Census and the Census on Establishments, in the grid-based data service menu. These datasets contain both \textit{hierarchical key variables} representing multiple grid levels (e.g., 100m, 1km, 10km, and 100km) as well as administrative divisions (e.g., province, city, county, and district) and survey-specific \textit{key variables}. For instance, demographic characteristics such as gender and age are used in population censuses, while other surveys include their own domain-specific attributes. The \texttt{iLBA} algorithm ensures confidentiality by controlling both disclosure risk and information loss during the aggregation of masked frequency tables and complements the Small Cell Adjustment technique used in the system.

Building upon this foundation, the present work introduces \textit{the first official and open-source implementation} of the iLBA algorithm as an \textsf{R} package. While the original Java version was tightly integrated within SGIS+, the \textsf{R} package makes the methodology broadly accessible to the global community of statistical agencies, researchers, and data providers. It offers reproducible and efficient tools for generating masked and aggregated frequency tables and assessing information loss. This implementation bridges theoretical development and practical application by enhancing the accessibility, transparency, and reproducibility of disclosure control methods for official statistics, allowing statistical offices to adopt the confidentiality-preserving approach used in SGIS+ for their own data dissemination systems.

\section*{Funding}
This work was supported by the National Research Foundation of Korea
(NRF) grants funded by the Korea government (MSIT) (RS-2024-00333399).


\newpage

\appendix

\section{Pitfalls of naive application of the SCA method}\label{pitfalls}
\setcounter{table}{0}
\renewcommand{\thetable}{A.\arabic{table}}

If one naively applies the SCA rule to the aggregated count of small frequency cells and releases
$\tilde{f}_{\Sc}^{\mathrm{SCA}}=6$, users can narrow down the possible true
counts of the small frequency cells in the finest level table. From Table~\ref{selected cells},
the released SCA-masked values imply that $f_j \in \{1,2,\dots,5\}$ for $j\in\{4,5,6,7\}$
and $f_j \in \{0,1,\dots,4\}$ for $j\in\{8\}$. Hence, the minimum feasible values are
$1$ for each cell in $\Sc_K$ and $0$ for each cell in $\Sc_0$, which sum to $4$.
The residual, $6-4=2$, must therefore be allocated across these cells. Table~\ref{feasible values} lists all feasible combinations, up to
permutation of $(f_4,f_5,f_6,f_7)$. It follows that each of $f_4,f_5,f_6,f_7$ lies in
$\{1,2,3\}$. Thus, the released value reveals that the cells in $\Sc_K$ are necessarily
small frequency cells smaller than $K$, which violates $K$-anonymity at the finest level. In contrast, no such conclusion can be drawn for $f_8$, because some feasible configurations allow $f_8=0$, which still satisfies $K$-anonymity.

\begin{table}[h]
\centering
\caption{Feasible combinations of $(f_4,f_5,f_6,f_7,f_8)$ consistent with
$\tilde{f}_{\Sc}^{\mathrm{SCA}}=6$, up to permutation of $(f_4,f_5,f_6,f_7)$}
\label{feasible values}
\begin{tabular}{lccccc}
\toprule
case & $f_4$ & $f_5$ & $f_6$ & $f_7$ & $f_8$ \\
\midrule
1 & 2 & 1 & 1 & 1 & 1 \\
2 & 2 & 2 & 1 & 1 & 0 \\
3 & 3 & 1 & 1 & 1 & 0 \\
4 & 1 & 1 & 1 & 1 & 2 \\
\bottomrule
\end{tabular}
\end{table}

\section{How K-ambiguity resolves differencing-based inference?}\label{K-ambiguity}

We take a closer look at when violation of $K$-anonymity at finest level during the aggregation occurs and generalize the situation. Let $D$ denote the interval of feasible values for $f_{\Sc}$ that users can infer from the released finest level table with the SCA rule, which given by
\begin{equation}
    D = \Big\{|\Sc_K|,|\Sc_{K}|+1,\dots,\; K|\Sc_K| + (K-1)|\Sc_0|\Big\}.
    \tag{B.1}
    \label{2}
\end{equation}
The lower bound is achieved by assigning the smallest feasible values to $f_j$, namely $f_j = 0$ for $j \in \Sc_0$ and $f_j = 1$ for $j \in \Sc_K$, whereas the upper bound is achieved by assigning the largest feasible values, namely $f_j = K-1$ for $j \in \Sc_0$ and $f_j = K$ for $j \in \Sc_K$.
Intuitively, a violation of $K$-anonymity at the finest level arises when the true total $f_{\Sc}$ lies so close to the boundary of this interval that the small frequency cells can be almost pinned down. In other words, to be safe from such inference, $f_{\Sc}$ should be at least $K-1$ away from either boundary point of $D$.

We first consider the case where $f_{\Sc}$ is close to the lower bound of $D$. 
A user attempting to infer the individual counts $f_j$ for $j \in \Sc$ would first assign the minimum feasible values consistent with the SCA rule, namely, $f_j = 0$ for $j \in \Sc_0$ and $f_j = 1$ for $j \in \Sc_K$. 
These assignments yield a baseline total of $|\Sc_K|$, which is the lower bound of $D$. 
The residual total
\begin{equation*}
    R = f_{\Sc} - |\Sc_K|
\end{equation*}
must then be allocated among the cells in $\Sc$, subject to $0 \le f_j \le K-1$ for $j \in \Sc_0$ and $1 \le f_j \le K$ for $j \in \Sc_K$. 
If $R \ge K-1$, then at least one cell in $\Sc_K$ can still attain frequency $K$ by assigning all $K-1$ residual units to that cell. 
Hence, a violation of $K$-anonymity at the finest level cannot yet be concluded.

In contrast, if $R < K-1$, then even if all the remaining amount is allocated to one cell, every $j \in \Sc_K$ satisfies $1 \le f_j \le K-1$. In this situation the users can conclude that no finest level cell in $\Sc_K$ reaches frequency $K$, and thus $K$-anonymity of $f_j,j\in \Sc_K$ is violated. Note that $K$-anonymity of $\Sc_0$ is not violated, since each cell $f_j, j\in \Sc_0$ can be assigned to be 0.

Next we consider the case where $f_\Sc$ is close to the upper boundary of $D$. To reason about this case, we start from the opposite extreme: assign the maximal feasible values to all finest level cells, that is, set $f_j = K$ for $j \in \Sc_K$ and $f_j = K-1$ for $j \in \Sc_0$. Denote $U=K|\Sc_K| + (K-1)|\Sc_0|$, as the upper bound of $D$. In order to reach the observed total $f_{\Sc}$, the users must subtract
\[
R^{\mathrm{up}} = U - f_\Sc
\]
from some of the cells while keeping every cell within its allowed range $[1,K]$ for $\Sc_K$ and $[0,K-1]$ for $\Sc_0$.

If $R^{\mathrm{up}} \ge K-1$, there is enough slack to reduce at least one cell in $\Sc_0$ from $K-1$ down to $0$ (subtracting $K-1$ from that cell) and then adjust the remaining cells, so a configuration with $f_j = 0$ for some $j \in \Sc_0$ is still possible. In this case the users cannot rule out that some finest level cell in $\Sc_0$ has true count $0$, and $K$-anonymity at the finest level may hold. 

In contrast, if $R^{\mathrm{up}} < K-1$, the total reduction $U - f_\Sc$ is not large enough to subtract $K-1$ from any cell in $\Sc_0$, so no cell in $\Sc_0$ can be reduced from $K-1$ to $0$. Hence every $j \in \Sc_0$ must satisfy $f_j \ge 1$. Together with the upper bound $f_j \le K-1$, this implies $1 \le f_j \le K-1 \quad \text{for all } j \in \Sc_0$ so all finest level cells in $\Sc_0$ are forced to be small but positive, which again violates $K$-anonymity of $f_j,j\in \Sc_0$.

To prevent such violations of $K$-anonymity when $f_\Sc$ lies close to the boundary of $D$, the
released information must leave sufficiently many feasible values for $f_\Sc$ inside $D$. By endowing $K$-ambiguity to $f_{\Sc}$,
the users can no longer almost uniquely determine any finest level count as a small positive value.

\section{Details of iLBA algorithm}\label{algorithm detail}

We first consider three cases of the set $\mathcal{S}$. Denote $\tilde{f}$ as masked aggregated count of $f$.
First, if there is \textbf{no small frequency cell} ($|\Sc|=0$), the aggregation consists only of large cells and no adjustment is required. In this case, $f = f_{\Lc}$ and hence 
\begin{equation}
    \tilde{f}= f_{\Lc}
\tag{C.1}
\label{3}
\end{equation}

Second, consider the case of \textbf{a single small frequency cell} ($|\Sc|=1$).  Let $j_0$ be its index so that $\Sc= \{j_0\}$. In this situation, we are allowed to release $\tilde f^{\mathrm{SCA}}_{j_0}$ which is obtained from the finest level table, since $K$-anonymity of it is ensured in both level. This situation is essentially equivalent to releasing the finest level table. The masked aggregated count is simply 
\begin{equation}
    \tilde{f} = f_{\Lc}+ \tilde f^{\mathrm{SCA}}_{j_0}
    \tag{C.2}
\label{4}
\end{equation}
Note that the SCA is applied only once to create the finest level table (i.e $\tilde{f}_j^{\text{SCA}}$) that are saved as a database in a system, and here we simply use these masked counts as given.
Thus, the masking procedure illustrated here involves no additional randomness from $\tilde{f}_j^{\text{SCA}}$.

The last case is when \textbf{multiple small frequency cells} are present ($|\Sc| \ge2$). The subcase $f_{\Sc} = 0$ implies that the aggregation consists only of zeros. Since applying the SCA method to zero leaves it unchanged, we can regard the aggregated count $f_{\Sc} = 0$ as already masked by the SCA, just as in (3). Hence, it remains to consider the nontrivial subcase $f_{\Sc}\ge 1$, for which the iLBA must be applied.

As discussed in Section~2, we introduce $K$-ambiguity into $f_S$ to guarantee $K$-anonymity at the finest level.
We endow $K$-ambiguity through the following first step:
\begin{equation}
  \tilde{f}_\Sc^{(1)} 
  = f_\Sc - \operatorname{mod}(f_\Sc - 1, K) + \big\lfloor K/2 \big\rfloor, 
  \tag{C.3}
  \label{5}
\end{equation}
where $\operatorname{mod}(a,K)$ is the remainder when $a$ is divided by $K$, and $\lfloor K/2 \rfloor$ is the greatest integer less than or equal to $K/2$.

From $\tilde{f}_\Sc^{(1)}$ in \eqref{5}, the users can infer that the true total $f_S$ lies in the following set of $K$ candidate values:
\begin{equation}
  C 
  = \Big\{
      \tilde{f}_\Sc^{(1)} - \big\lfloor K/2 \big\rfloor,\,
      \dots,\,
      \tilde{f}_\Sc^{(1)} - \big\lfloor K/2 \big\rfloor + K - 1
    \Big\}.                                                 \tag{C.4}
    \label{6}
\end{equation}

However, some of these candidates may partly lie outside the feasible interval
$D$ defined in \eqref{2}. In such a case, only the portion $C \cap D$ is
inside $D$, and it may contain fewer than $K$ feasible candidates, which breaks
$K$-anonymity at the finest level. To prevent this, we adjust
$\tilde{f}^{(1)}_\Sc$ so that every candidate in interval $C$ is entirely contained
in $D$.

We can observe that, given $K \ge 2$,  we have $|C| = K \le |D|$, so the length of $D$ is always at least
as large as that of $C$. Hence $C$ can never fully cover $D$.
Intuitively, when the range of $C$ is not fully contained in the range of $D$, the range of $C$ is partly lie outside the range of $D$ on at most one side. If we denote the lower and upper boundaries of
an interval $I$ by $\min(I)$ and $\max(I)$, respectively, then the range of $C$
is not fully contained in the range of $D$ precisely when
\begin{equation}
    \min(C) < \min(D)
    \quad \text{or} \quad
    \max(D) < \max(C).
    \tag{C.5}
\end{equation}

Under this condition, the two sets $C$ and $D$ still overlap. We now show that
we can always move $C$ into $D$ by shifting it by one block of size $K$.

Consider the case $\min(C) < \min(D)$. We shift $C$ by $K$ units to the right
and define $C' = C + K$. From the explicit forms of $C$ and $D$ in \eqref{2},\eqref{6}, a
simple calculation shows that
\[
  \min(D) \le \min(C') \quad \text{and} \quad \max(C') \le \max(D),
\]
hence $C' \subset D$. The other case $\max(D) < \max(C)$ is symmetric and is
handled by shifting $C$ to the left by $K$.

Hence, by adding or subtracting $K$ from $\tilde{f}_\Sc^{(1)}$, we can shift the entire range $C$ into $D$ while keeping its length equal to $K$.  
Formally, we define
\begin{equation}
  \tilde{f}_\Sc^{(2)} =
  \begin{cases}
    \tilde{f}_\Sc^{(1)} + K, & \text{if } \min(C) < \min(D),\\[4pt]
    \tilde{f}_\Sc^{(1)} - K, & \text{if } \max(D) < \max(C),\\[4pt]
    \tilde{f}_\Sc^{(1)},     & \text{otherwise.}
  \end{cases}
  \tag{C.6}
  \label{8}
\end{equation}
Note that when the shift occurs in the case where $\min(C)<\min(D)$, it is refered to as type1, whereas the other case is referred to as type2.
This step produces a new candidate set $C'$ of size $K$ that lies entirely within $D$, thereby preserving $K$-ambiguity.

To avoid releasing ambiguously masked values that are strictly between $0$ and
$K$, i.e. in the range $\{1,\dots,K-1\}$, iLBA applies a final post–processing
rule. From (5)–(8), we have
\[
  \tilde f^{(2)}_\Sc = qK + 1 + \bigl\lfloor K/2\bigr\rfloor
\]
for some integer $q\ge0$, so the only possible value of $\tilde f^{(2)}_\Sc$
strictly between $0$ and $K$ is $1 + \lfloor K/2\rfloor$. We therefore define
\begin{equation}
  \tilde f^{\text{(3)}}_\Sc =
  \begin{cases}
    K, & \tilde f^{(2)}_\Sc = 1 + \bigl\lfloor K/2\bigr\rfloor,\\[0.2em]
    \tilde f^{(2)}_\Sc, & \text{otherwise}.
  \end{cases}
  \tag{C.7}
  \label{9}
\end{equation}
Thus, the released value is either $0$ or at least $K$.

\vspace{0.5em}
The iLBA algorithm is summarized as:
\begin{equation}
  \tilde f^{\text{iLBA}}_{\Sc} =
  \begin{cases}
    0, & f_{\Sc}=0  \quad \text{or} \quad |\Sc|=0,\\[0.2em]
    \tilde f^{\mathrm{SCA}}_{j_0}, & \Sc=\{j_0\},\\[0.2em]
    f^{\text{(3)}}_\Sc, & |\Sc|\geq2, f_{\Sc}\geq1.
  \end{cases}
  \tag{C.8}
  \label{10}
\end{equation}
Here, $\tilde{f}^{\text{SCA}}_{j0} \in \{0, K\}$. 
Moreover, $\tilde{f}^{(3)}_\Sc$ equals either $K$ or $qK + 1 + \lfloor K/2\rfloor$ for some integer $q \geq 1$.

\end{document}